\documentstyle[12pt,a4,cite,rotate,epsf]{article}
\newcommand{\cD}{{\cal D}}
\newcommand{\wh}{\widehat}
\newcommand{\ep}{\epsilon}

\newcommand{\nn}{\nonumber}

\newcommand{\fm}{\mbox{\rm fm}}

\newcommand{\eqn}[1]{(\ref{#1})}

\newcommand{\mev}{\mbox{\rm MeV}}
\newcommand{\gev}{\mbox{\rm GeV}}

\newcommand{\tvs}{\vbox{\vskip 6mm}}
\newcommand{\cDt}{\widetilde{\cal D}}
\newcommand{\smvs}{\vbox{\vskip 8mm}}

\newcommand{\newsection}[1]{\section{#1}\setcounter{equation}{0}}

\newcommand{\vacl}{\langle 0|}
\newcommand{\vacr}{|0\rangle }
\newcommand{\prp}{\frac{1}{2}\,(1+\!\!\not{\!v})}
\newcommand{\aFF}{\langle aFF\rangle}
\newcommand{\expo}{e^{\,gf^{abc}z^\tau \int_0^1 dt A^c_\tau(x+tz)}}

\begin{document}
%\bibliographystyle{physics}

%%%%%%%%%%%%%%%%%%%%%%%%%%%%%%%%%%%%%%%%%%%%%%%%%%%%%%%%%%%%%%%%%%%%%%%%%
% The title page
%%%%%%%%%%%%%%%%%%%%%%%%%%%%%%%%%%%%%%%%%%%%%%%%%%%%%%%%%%%%%%%%%%%%%%%%%

\date{\small (December 1998)}

\author{
{\normalsize\bf H.~G.~Dosch, M.~Eidem\"uller and M.~Jamin\footnote{Heisenberg
 fellow} } \\
\ \\
{\small\sl Institut f\"ur Theoretische Physik, Universit\"at Heidelberg,} \\
{\small\sl Philosophenweg 16, D-69120 Heidelberg, Germany}\\
}

\title{
{\small\sf
\rightline{HD-THEP-98-51}
\rightline{hep-ph/9812417}
}
\bigskip
\bigskip
{\Huge\bf QCD sum rule analysis of \\
 the field strength correlator \\}
}

\maketitle
\thispagestyle{empty}

\begin{abstract}
\noindent
The gauge invariant two-point correlator for the gluon field strength tensor
is analysed by means of the QCD sum rule method. To this end, we make use of
a relation of this correlator to a two-point function for a quark-gluon
hybrid in the limit of the quark mass going to infinity. From the sum rules
a relation between the gluon correlation length and the gluon condensate is
obtained. We briefly compare our results to recent determinations of the
field strength correlator on the lattice.
\end{abstract}

\vspace{1cm}


\newpage
\setcounter{page}{1}

%%%%%%%%%%%%%%%%%%%%%%%%%%%%%%
% The main part of the paper %
%%%%%%%%%%%%%%%%%%%%%%%%%%%%%%

%%%%%%%%%%%%%%%%
% Introduction %
%%%%%%%%%%%%%%%%

\newsection{Introduction}

The gauge invariant non-local gluon field strength correlator plays an
important r\^ole in non-perturbative approaches to QCD \cite{svz:79,vol:79,
leu:81,sim:88,dos:94}. It is the basic ingredient in the model of the
stochastic vacuum (MSV) \cite{dos:87,ds:88} and in the description of high
energy hadron-hadron scattering \cite{nr:84,ln:87,kd:90,dfk:94}. In the
spectrum of heavy quark bound states it governs the effect of the gluon
condensate on the level splittings \cite{gro:82,cgo:86,kdb:92,sty:95} and
it is useful for the determination of the spin dependent parts in the
heavy quark potential \cite{sd:88,sim:89}.

Its next-to-leading order correction in perturbative QCD has been calculated
recently by two of the authors \cite{ej:98}. The correlator has also been
measured on the lattice in pure gauge theory and full QCD using the cooling
method \cite{gmp:97,egm:97} and by making the assumptions of the MSV from
lattice calculations of the heavy quark potentials \cite{bbv:98}. The lattice
analyses found that for distances $z$ of the gluon field strength larger than
roughly $0.4\,{\rm fm}$ an exponential decaying term dominates yielding a
correlation length of approximately $0.2\,{\rm fm}$. On the other hand the
short distances are dominated by the perturbative $1/z^4$ behaviour. Recently,
the field strength correlator has also been calculated in the framework of
exact renormalisation group equations \cite{elw:98}.

The gauge invariant gluon field strength correlator can be related to
a correlator of a colour singlet current composed of a (fictitious)
infinitely heavy octet quark and the gluon field strength tensor. This
fact has already been employed in ref. \cite{ej:98} in order to apply the
machinery developed in the Heavy Quark Effective Theory\footnote{For a
review on HQET as well as original references the reader is referred to
\cite{neu:94}.} (HQET) for calculating the perturbative corrections. In this
paper we again use this relation in order to apply QCD sum rule techniques \cite{svz:79}
to the correlator in question. The sum rule analysis can be used to estimate
the correlation length of the field strength correlator using as ingredients
the value of the gluon condensate and the results for the perturbative
calculation.

Our paper is organised as follows. In the next section we discuss again the
relation of the field strength correlator and the corresponding heavy quark
current correlator. In section~3 we set up the different contributions
needed for the sum rule analysis and in section~4 we present our results
together with a comparison with recent lattice determinations of the
field strength correlator. Finally, in section~5, we end with some conclusions
and an outlook.

%%%%%%%%%%%%%%%%%%%%%%%%%%%%%%%%%
% The field strength correlator %
%%%%%%%%%%%%%%%%%%%%%%%%%%%%%%%%%

\newsection{The field strength correlator}

The gauge invariant two-point correlation function of the QCD field strength
tensor $F^a_{\mu\nu}(x)$ in the adjoint representation can be defined as
\begin{equation}
\label{eq:2.1}
\cD_{\mu\nu\rho\sigma}(z) \; \equiv \; \langle 0|T\{g_s^2 F^a_{\mu\nu}(y)
{\cal P}\expo F^b_{\rho\sigma}(x)\}|0\rangle \,,
\end{equation}
where the field strength $F^a_{\mu\nu}=\partial_\mu A^a_\nu-\partial_\nu
A^a_\mu+gf^{abc}A^b_\mu A^c_\nu$, $z=y-x$ and ${\cal P}$ denotes path
ordering of the exponential. In general, the gauge invariant field strength
correlator could be defined with an arbitrary gauge string connecting the
end points $x$ and $y$, but in this work we shall restrict ourselves to
a straight line. Only for that case the relation to HQET is possible.
From the Lorentz structure of the field strength correlator it follows
that the correlator can be parametrised in terms of two scalar functions
$\cD(z^2)$ and $\cD_1(z^2)$ \cite{ds:88}:
\begin{eqnarray}
\label{eq:2.2}
\cD_{\mu\nu\rho\sigma}(z) & = & \Big[\,g_{\mu\rho}g_{\nu\sigma}-
g_{\mu\sigma}g_{\nu\rho}\,\Big]\Big(\,\cD(z^2)+\cD_1(z^2)\,\Big) \nn \\
\smvs
& & \hspace{-3.9mm} +\,\Big[\,g_{\mu\rho}z_\nu z_\sigma-g_{\mu\sigma}
z_\nu z_\rho-g_{\nu\rho} z_\mu z_\sigma+g_{\nu\sigma}z_\mu z_\rho
\,\Big]\,\frac{\partial\cD_1(z^2)}{\partial z^2} \,.
\end{eqnarray}
The invariant function $\cD(z^2)$ can only occur in a non-abelian gauge
theory or an abelian one with monopoles. In the MSV it is responsible for
confinement and the formation of a string.

The correlator $\cD_{\mu\nu\rho\sigma}(z)$ can be related to the correlator
of a local, gauge invariant current composed of an infinitely heavy quark
field in the octet representation, $h^a(x)$, and the gluon field strength
tensor \cite{eid:97,ej:98}. The current in question takes the form
$(g_s h^a F^a_{\mu\nu})(x)$. Analogously to HQET the heavy octet-quark field
is constructed from the field $Q^a$ with a finite mass $m_Q$ in the limit
\begin{equation}
\label{eq:2.3}
h^a(x) \; = \; \lim_{m_Q\rightarrow\infty}\,\prp\,e^{im_Qvx}Q^a(x) \,,
\end{equation}
with $v$ being the four-velocity of the heavy quark. 
The propagator of the free heavy quark field $h^a_0(x)$ in coordinate
space is given by
\begin{equation}
\label{eq:2.4}
S(z) \; = \; \vacl T\{h^a_0(y)\bar{h}^b_0(x)\} \vacr \; = \; \delta^{ab}
\frac{1}{v^0}\,\theta(z^0)\,\delta\Big({\bf z}-\frac{z^0}{v^0}{\bf v}\Big) \,,
\end{equation}
where  $v^0$ is the zero-component of the velocity.
The correlator of the full field can be obtained by integrating out only
the heavy quark and leaving the expectation value with respect to the
gauge field:
\begin{equation}
\label{eq:2.5}
\vacl T\{h^a(y)\bar{h}^b(x)\} \vacr \; = \;
 S(z)\,\vacl {\cal P}\expo \vacr \,.
\end{equation}
The gauge string is left after the elimination of the heavy quarks from the
interaction term of adjoint quarks with the colour potential
\begin{equation}
\label{eq:2.6}
{\cal L}_{int} \; = \; -\,ig_s f_{abc} v^\mu \bar{h}^a(x) A_\mu^c(x) h^b(x) \,.
\end{equation}
The physical picture of this result is a heavy quark moving from point
$x$ to $y$ with a four-velocity $v$, acquiring a phase proportional to the
path-ordered exponential. 
The limit of $m_Q\rightarrow\infty$ is necessary in order to constrain the
heavy quark on a straight line and in order to decouple the spin interactions.
The same relation also holds for quarks in the fundamental representation
with the appropriate replacements in the exponential.

The equation \eqn{eq:2.5} allows to establish a relation between the field
strength correlator \eqn{eq:2.1} and the correlator for the colourless
heavy quark current.
By integrating out the heavy degrees of freedom and using \eqn{eq:2.5} we
arrive at
\begin{eqnarray}
\label{eq:2.8}
\cDt_{\mu\nu\rho\sigma}(z)
& \equiv & \vacl T\{g_s^2 F^a_{\mu\nu}(y)h^a(y)F^b_{\rho\sigma}(x)\bar{h}^b(x)
\}\vacr \nn \\
& = & S(z) \, D_{\mu\nu\rho\sigma}(z) \,.
\end{eqnarray}
We may view the composite operator $(g_s h^a F^a_{\mu\nu})(x)$ as an
interpolating field of colourless quark gluon hybrids and evaluate
$\cDt_{\mu\nu\rho\sigma}(z)$ by introducing these as intermediate states
in the absorption part of $\cDt_{\mu\nu\rho\sigma}(z)$. The lowest lying
state will govern the long-range behaviour and hence the inverse of its
energy is the correlation length.

Our next aim is to evaluate this correlator in the framework of QCD sum rules
\cite{svz:79} and in that way obtain information on the correlation length
of the gluon field strength correlator.
For the sum rule analysis it is preferable to work with the correlator in
momentum space. Thus we define
\begin{equation}
\label{eq:2.9}
\cDt_{\mu\nu\rho\sigma}(w) \; = \; i \int dz \, e^{iqz} \vacl
T\{g_s^2 F^a_{\mu\nu}(y)h^a(y) F^b_{\rho\sigma}(x)\bar{h}^b(x)\}\vacr \,,
\end{equation}
where the residual heavy quark momentum is $w=vq$. Similar to the Lorentz
decomposition of the coordinate space correlator $\cD_{\mu\nu\rho\sigma}(z)$
into scalar functions $\cD(z^2)$ and $\cD_1(z^2)$, eq.~\eqn{eq:2.2},
we can write the momentum space correlator as follows: 
\begin{eqnarray}
\label{eq:2.10}
\cDt_{\mu\nu\rho\sigma}(w) & = & \Big[\,g_{\mu\rho}g_{\nu\sigma}-
g_{\mu\sigma}g_{\nu\rho}\,\Big]\Big(\,\cDt(w)+\cDt_1(w)\,\Big) \nn \\
\smvs
& & \hspace{-3.9mm} +\,\Big[\,g_{\mu\rho}v_\nu v_\sigma-g_{\mu\sigma}
v_\nu v_\rho-g_{\nu\rho} v_\mu v_\sigma+g_{\nu\sigma}v_\mu v_\rho
\,\Big]\,\cDt_*(w) \,.
\end{eqnarray}
The functions $\cDt(w)$ and $\cDt_1(w)$ are the Fourier transforms of
$S(z)\,\cD(z^2)$ and $S(z)\,\cD_1(z^2)$ respectively, the function
$\cDt_*(w)$ is the Fourier transform of
$S(z)z^2\partial \cD_1(z^2)/\partial z^2$.

For our purpose of isolating intermediate states of the correlator
\eqn{eq:2.9} a decomposition according to an $O_3$ classification is more
appropriate than the decomposition of eq.~\eqn{eq:2.10}. Since the spin of
the heavy quark decouples, we only have to consider the gluon spin. The
six-component field can be decomposed into tensor structures depending on the
only two external vectors in the game; the four-velocity $v_\mu$ and the
polarisation vector of the gluon $e_\mu$. This leads to the two Lorentz
structures for the hadronic matrix elements
\begin{eqnarray}
\label{eq:2.11}
\vacl (g_s F^a_{\mu\nu} h^a)(0)| H^{-}\rangle & = &
f^-\,(v_\mu e_\nu-v_\nu e_\mu) \,, \\
\tvs
\vacl (g_s F^a_{\mu\nu} h^a)(0)| H^{+}\rangle & = &
f^+\,\varepsilon_{\mu\nu\lambda\kappa}v^\lambda e^\kappa \,,
\end{eqnarray}
where $H^\mp$ are hadronic states with the same quantum numbers as the
composite current. In the rest frame ${\bf v}=0$, the first structure
transforms as a 3-vector and thus $H^-$ corresponds to a $1^-$ state whereas
the second structure transforms as an axialvector and $H^+$ corresponds to a
$1^+$ state.

Through appropriate projections the two quantum numbers can be singled
out from the correlator $\cDt_{\mu\nu\rho\sigma}(w)$. Hence, we define
\begin{eqnarray}
\label{eq:2.13}
\cDt^-(w) & \equiv & g^{\mu\rho}v^\nu v^\sigma \, \cDt_{\mu\nu\rho\sigma}(w)
\; = \; 3\,\Big(\cDt(w)+\cDt_1(w)+\cDt_*(w)\Big) \,, \\
\tvs
\label{eq:2.14}
\cDt^+(w) & \equiv & (g^{\mu\rho}g^{\nu\sigma}-2\,g^{\mu\rho}v^\nu v^\sigma) \,
\cDt_{\mu\nu\rho\sigma}(w) \; = \; 6\,\Big(\cDt(w)+\cDt_1(w)\Big) \,.
\end{eqnarray}
The Fourier transforms of the functions $\cDt^-(w)$ and $\cDt^+(w)$ are
up to the factor $S(z)$ the invariant functions $\cD_\parallel(z^2)$ and
$\cD_\perp(z^2)$ respectively which have been used in the lattice
calculations of refs.~\cite{gmp:97,egm:97}.

Under the assumption of quark-hadron duality which is usually made for
sum rule analyses \cite{svz:79}, we model the correlators by
a contribution from the lowest lying resonances plus the perturbative
continuum above a threshold $s_0$. Inserting the matrix elements and
performing the heavy quark phase space integrals one obtains
\begin{equation}
\label{eq:2.15}
\cDt^\mp(w) \; = \; \frac{\kappa^\mp\,|f^\mp|^2}{w-E^\mp+i\ep} +
\int\limits_{s_0^\mp}^\infty d\lambda \, \frac{\rho^\mp(\lambda)}
{\lambda-w-i\ep} \,,
\end{equation}
where $\kappa^-\!=1$, $\kappa^+\!=-2$ and $E$ represents the energy
of the glue around the heavy quark. The spectral densities are defined by
$\rho^\mp(\lambda)\equiv 1/\pi\,{\rm Im}\,\cDt^\mp(\lambda+i\ep)$ and are
known at the next-to-leading order \cite{ej:98}. Explicit expressions will
be given in the next section.

After Fourier transformation to coordinate space the above representation
reads:
\begin{eqnarray}
\label{eq:2.16}
\cDt(z) & = & -i \int \frac{d^4 q}{(2 \pi)^4} \, e^{-iqz}\,\cDt(w) \nn \\
& = & \biggl\{-\kappa\,|f|^2 e^{-iE|z|} + \int\limits_{s_0}^\infty d\lambda \,
\rho(\lambda) \, e^{-i\lambda |z|} \biggr\} \, S(z) \,,
\end{eqnarray}
where the factorisation of the heavy quark propagator can be seen explicitely.
The inverse correlation length is found to be given by $E$.

%%%%%%%%%%%%%%%%%
% The sum rules %
%%%%%%%%%%%%%%%%%

\newsection{The sum rules}

The {\em phenomenological side} of the sum rules has already been given by
eq.~\eqn{eq:2.15}. In this section, we shall present the {\em theoretical side}
of the sum rules which arises from calculating the correlator of
eq.~\eqn{eq:2.9} in the framework of the operator product expansion
\cite{svz:79,wil:69}.

In coordinate space the purely perturbative contribution up to the
next-to-leading order in the strong coupling constant has been calculated
in ref.~\cite{ej:98}. Here we give the corresponding results in momentum
space for $\cDt^\mp(w)$:
\begin{equation}
\label{eq:3.1}
\cDt^\mp_{PT}(w) \; = \; (-w)^3\, a\,\Big[\, p_{10}^\mp+p_{11}^\mp L +
a\, (p_{20}^\mp+p_{21}^\mp L+p_{22}^\mp L^2) \,\Big] \,,
\end{equation}
where $a\equiv\alpha_s/\pi$, $L=\ln(-2w/\mu)$ and the coefficients
$p_{ij}^\mp$ are given explicitly in the appendix. From this result one can
immediately calculate the corresponding spectral functions:
\begin{equation}
\label{eq:3.2}
\rho^\mp(\lambda) \; = \; \lambda^3\,a\, \biggl[\, p_{11}^\mp +
a \,\biggl(p_{21}^\mp +  2\,p_{22}^\mp\ln \frac{2 \lambda}{\mu}\biggr)
\,\biggr] \,,
\end{equation}
where $\lambda$ has to be greater zero. Essential for the sum rule analysis
are the contributions coming from
the condensates. The correlation function is expanded in powers of $1/w$
corresponding to higher and higher dimensional condensates. In our case
the dimension three condensate $\langle\bar{h}h\rangle$ vanishes since
the quark mass is infinite. The lowest nonvanishing term is the gluon
condensate of dimension four:
\begin{equation}
\label{eq:3.3}
\cDt^-_{FF}(w) \; = \; -\,\frac{\pi^2}{w}\aFF \,, \qquad
\cDt^+_{FF}(w) \; = \; -\,\frac{2\pi^2}{w}\aFF \,.
\end{equation}
The next condensate contribution would be of dimension six, but we shall
neglect all higher condensate contributions in this work and restrict
ourselves to the gluon condensate.

In order to suppress contributions in the dispersion integral coming from
higher exited states and from higher dimensional condensates, it is convenient
to apply a Borel transformation $\wh{B}_T$ with $T$ being the Borel variable
\cite{svz:79}. Some useful formulae for the Borel transformation are also
collected in the appendix. For the phenomenological side of the sum rules,
eq.~\eqn{eq:2.15}, we then find
\begin{equation}
\label{eq:3.4}
\wh{\cD}^\mp(T) \; = \; -\,\kappa^\mp\,|f^\mp|^2 \,e^{-E^\mp/T} +
\int\limits_{s_0^\mp}^\infty d\lambda\,\rho^\mp(\lambda)\,e^{-\lambda/T} \,.
\end{equation}
For the perturbative contribution it is convenient to apply the following
identity:
\begin{equation}
\label{eq:3.5}
\wh{B}_T \,\cDt(w) \; = \; T^4 \,\wh{B}_T \left(\frac{d}{dw}\right)^{\!4}
\cDt(w) \,,
\end{equation}
from which we obtain
\begin{equation}
\label{eq:3.6}
\wh{\cD}^\mp_{PT}(T) \; = \; 6\, T^4\, a\,\biggl[\, p_{11}^\mp + a \,\biggl(
p_{21}^\mp+ \frac{1}{3}\,\Gamma'(4)\, p_{22}^\mp +  2\,p_{22}^\mp \ln
\frac{2T}{\mu}\biggr) \,\biggr] \,,
\end{equation}
where $\gamma_E$ is Eulers constant and $\Gamma'(4)=11-6\gamma_E$.
The Borel transformed expression for the gluon condensate contribution
is found to be:
\begin{equation}
\label{eq:3.7}
\wh{\cD}^-_{FF} \; = \;   \pi^2 \aFF \,, \qquad
\wh{\cD}^+_{FF} \; = \; 2 \pi^2 \aFF \,.
\end{equation}
After Borel transformation, the correlators satisfy homogeneous
renormalisation group equations. Thus we can improve the perturbative
expressions by resumming the logarithmic contributions. The perturbative
contribution is then expressed in terms of the running coupling
$a(2T)$:
\begin{equation}
\label{eq:3.8}
\wh{\cD}^\mp_{PT}(T) \; = \; 6\, T^4 \left(\frac{a(2T)}{a(\mu)}\right)^
{-\gamma_1^\mp/\beta_1} \!a(2T)\,\left[\, p_{11}^\mp + a\,\left(\,p_{21}^\mp
+\frac{1}{3}\,\Gamma'(4)\,p_{22}^\mp\right) \,\right] \,,
\end{equation}
where $\beta_1=11/2-n_f/3$ is the first coefficient of the QCD
$\beta$-function. Reexpanding and comparing with eq.~\eqn{eq:3.6},
the anomalous dimensions $\gamma_1^\mp$ are found to be
$\gamma_1^\mp=2\,p_{22}^\mp/p_{11}^\mp+\beta_1$, or explicitly
\begin{equation}
\label{eq:3.9}
\gamma_1^- \; = \; 0 \,, \qquad
\gamma_1^+ \; = \; 3 \,.
\end{equation}
Let us note that the correlator $\wh{\cD}^-(T)$ which corresponds to the
vector intermediate state does not depend on the renormalisation scale $\mu$
at this order.

For the continuum contribution we first evaluate the integral with the
general formula \cite{jm:95} which makes the numerical analysis easier:
\begin{equation}
\label{eq:3.10}
\int\limits_{s_0}^\infty d\lambda\,\lambda^{\alpha-1}\ln^n \frac{2\lambda}{\mu}
e^{-\lambda/T} \; = \; T^\alpha \sum_{k=0}^{n} {n\choose k} \ln^k\frac{2T}{\mu}
\left[\frac{\partial^{n-k}}{\partial\alpha^{n-k}}\Gamma\left(\alpha,\frac{s_0}
{T}\right)\right] \,,
\end{equation}
some formulae for the incomplete $\Gamma$-function $\Gamma(\alpha,x)$ are
given in the appendix. We then obtain
\begin{eqnarray}
\label{eq:3.11}
\chi^\mp(T,s_0) & = & \int\limits_{s_0}^\infty d\lambda\,\rho^\mp(\lambda) \,
e^{-\lambda/T} \; = \; T^4\,a\,\Biggl\{\,p_{11}^\mp \,\Gamma\left(4,
\frac{s_0}{T}\right) \\
\smvs
& & +\,a\left[\,\left(\,p_{21}^\mp+2\,p_{22}^\mp\ln\frac{2T}
{\mu}\right)\Gamma\left(4,\frac{s_0}{T}\right) + 2\,p_{22}^\mp\Gamma'
\left(4,\frac{s_0}{T}\right)\,\right]\,\Biggr\} \,, \nn
\end{eqnarray}
and after renormalisation group improvement
\begin{eqnarray}
\label{eq:3.12}
\chi^\mp(T,s_0) & = & T^4 \left(\frac{a(2T)}{a(\mu)}\right)^{-\gamma_1^\mp/
\beta_1}\!a(2T)\,\Biggl\{\,p_{11}^\mp \,\Gamma\left(4,\frac{s_0}{T}\right)
\hspace{30mm} \nn \\
\smvs
& & +\;a \left[\,p_{21}^\mp\,\Gamma\left(4,\frac{s_0}{T}\right)+2\,p_{22}^\mp
\,\Gamma'\left(4,\frac{s_0}{T}\right)\,\right]\,\Biggr\} \,.
\end{eqnarray}
In the limit $s_0\rightarrow 0$, eq.~\eqn{eq:3.12} agrees with eq.
\eqn{eq:3.8} as it should.

%%%%%%%%%%%%%%%%%%%%%%
% Numerical analysis %
%%%%%%%%%%%%%%%%%%%%%%

\newsection{Numerical analysis}

After equating the phenomenological and the theoretical part we end up with
the sum rule
\begin{equation}
\label{eq:4.1}
K^\mp(T) \; \equiv \; -\,\kappa^\mp |f^\mp|^2 e^{-E^\mp/T} \; = \;
\wh\cD_{FF}^\mp + \wh\cD_{PT}^\mp(T)-\chi^\mp(T,s_0) \,,
\end{equation}
where $\kappa^-= 1$ and $\kappa^+=-2$. In order to estimate the binding
energy we derive as an immediate consequence of \eqn{eq:4.1}:
\begin{equation}
\label{eq:4.2}
E^\mp \; = \; -\,\frac{\partial}{\partial(1/T)}\,\ln K^\mp \; = \;
-\,\frac{\frac{\partial}{\partial(1/T)}\Big(\wh\cD_{PT}^\mp(T)-\chi^\mp(T,s_0)
\Big)}{\Big(\wh\cD_{FF}^\mp + \wh\cD_{PT}^\mp(T)-\chi^\mp(T,s_0)\Big)} \,.
\end{equation}
The derivative can also be given analytically if we first derive with respect
to $T$ and then perform the resummation of the logarithms. We thus find
\begin{eqnarray}
\label{eq:4.3}
\lefteqn{\frac{\partial}{\partial (1/T)} \left(\wh\cD^\mp_{PT}(T) -
\chi^\mp(T,s_0)\right) \; =} \nn \\
& & -\,T^5 \left(\frac{a(2T)}{a(\mu)}\right)^{-\gamma_1^\mp/\beta_1}\!\!
a(2T)\,\Bigg\{\, p_{11}^\mp \left(\Gamma(5)-\Gamma\left(5,\frac{s_0}{T}\right)
\right) \nn \\
& & +\,a\,\Bigg[\,p_{21}^\mp \left(\Gamma(5)-\Gamma\left(5,\frac{s_0}{T}\right)
\right) +  2\,p_{22}^\mp \left(\Gamma'(5)-\Gamma'\left(5,\frac{s_0}{T}\right)
\right)\Bigg]\Bigg\} \,.
\end{eqnarray}
We note that the different signs of the perturbative and non-perturbative
terms in the $1^-$ state lead to a stabilisation for the energy sum rule,
whereas the equal sign in the $1^+$ state destabilises the sum rule.

Let us begin our numerical analysis with the case for three light quark
flavours. As our input parameters we use $\aFF = 0.024 \pm 0.012 \,\gev^4$
and $\Lambda_{3fl} = 325 \,\mev$. In principle, the coupling constant at
next-to-leading order could be evaluated at any scale $\mu$. As our central
value in the numerical analysis we have chosen $\mu = 2 \,\gev$.
For the energy $E^-$ of the $1^-$ state we obtain the
best stability for a continuum threshold $s_0 = 1.7 \,\gev$ in the range
$T \geq 0.7 \,\gev$ with an energy $E^- \approx 1.4 \,\gev$. To estimate
the errors we have varied the scale $\mu$ as well as the continuum
threshold $s_0$. In figure~1 we have displayed the energy $E^-$ as a
function of the Borel parameter $T$ for $\mu=1\,\gev$ (dashed lines),
$2\,\gev$ (solid lines) and $4\,\gev$ (dotted lines). The corresponding
values of the continuum threshold are $s_0=1.5\pm0.2\,\gev$, $1.7\pm0.2\,\gev$
and $1.9\pm0.2\,\gev$ respectively. The central values have been chosen
in order to obtain maximal stability for the sum rule.

\begin{figure}[thb]
\vspace{0.2cm}
\centerline{
\rotate[r]{
\epsfysize=13cm
\epsffile{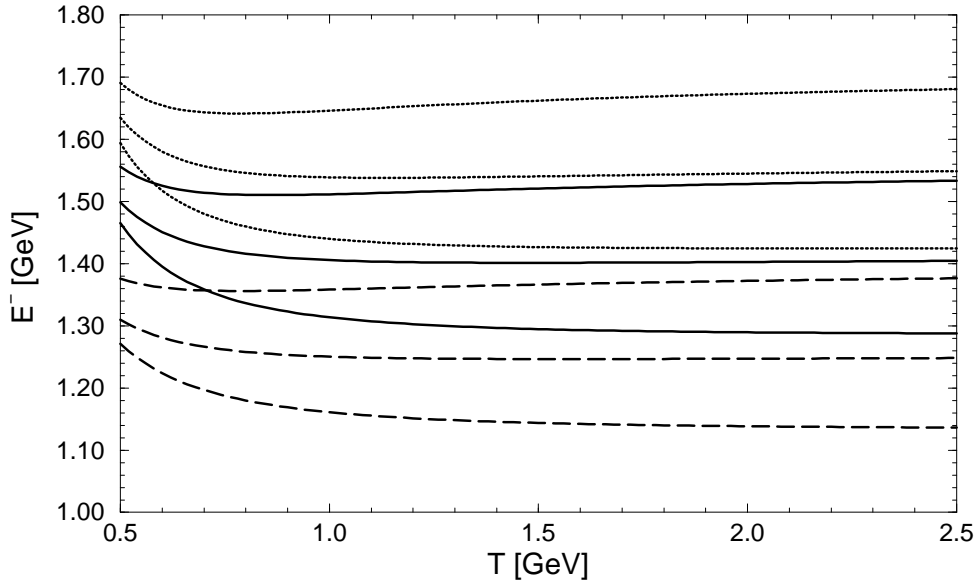} }}
\caption[]{The energy $E^-$ as a function of the Borel-parameter $T$
for three different renormalisation scales $\mu$ and continuum
thresholds $s_0$.
Dashed curves $\mu=1\,\gev$: lowest $s_0=1.3\,\gev$, middle $s_0=1.5\,\gev$,
upper $s_0=1.7\,\gev$.
Solid curves $\mu=2\,\gev$: lowest $s_0=1.5\,\gev$, middle $s_0=1.7\,\gev$,
upper $s_0=1.9\,\gev$.
Dotted curves $\mu=4\,\gev$: lowest $s_0=1.7\,\gev$, middle $s_0=1.9\,\gev$,
upper $s_0=2.1\,\gev$.
\label{fig:1}}
\end{figure}

Larger values of $s_0$ always increase $E$ but at the same time the stability
region shrinks and goes to smaller values of $T$. However, even at
$T = 0.7 \,\gev$ the influence of the higher resonances expressed through
the continuum model  $\chi^-(T)$ is very large:
$\chi^-(0.7)/\cD^-_{PT}(0.7) \approx 0.75$. For $s_0 = 2.1 \,\gev$, we have
a small stability region around $T = 0.65 \,\gev$
yielding $E^- = 1.6 \,\gev$. Here the influence of the continuum model is
around tolerable 50\%, the perturbative corrections and the choice of the
renormalisation scale become however more important there.

Another source of uncertainty is the value of the gluon condensate. For the
value $\aFF = 0.012 \,\gev^4$, originally obtained by \cite{svz:79}, we
find $E^- = 1.2 \,\gev$ at $s_0 = 1.5 \,\gev$, whereas for
$\aFF = 0.036 \,\gev^4$ we obtain $E^- = 1.8 \,\gev$ at $s_0 = 2.4 \,\gev$. 
We therefore conclude from the sum rules for the above mentioned parameters
an energy $E^-$ and a correlation length $a^-$ of:
\begin{equation}
E^-_{3fl} \; = \; 1.5 \pm 0.4 \;\gev \qquad {\rm and} \qquad
a^-_{3fl} \; = \; 0.13^{+0.05}_{-0.02} \;\fm \,.
\end{equation}

The main sources of uncertainty are the value of the gluon condensate and the
continuum contribution. Though the perturbative two-loop contributions to
the sum rule are very large, their influence on the value of $E^-$ is not so
dramatic. The corrections tend to cancel in the ratio of eq. \eqn{eq:4.2}.
If one determines the energy from the sum rule just containing the lowest
order perturbation theory and chooses as the scale for $\alpha_s$ the
approximate value of the energy one finds for $\aFF = 0.024 \,\gev^4$ the
value $E^- = 1.9 \,\gev$.

In a world without light quarks, i.e. $n_f = 0$, the main influence on the
sum rule is the expected change of the gluon condensate which might increase
by a factor two to three \cite{nsvz:84}. If we perform an analysis as above,
we get for $\Lambda_{0fl} = 250 \,\mev$ \cite{lue:98},
$\aFF = 0.048 \pm 0.024 \,\gev^4$ and $s_0 = 2.3 \,\gev$ an energy and
correlation length of 
\begin{equation}
E^-_{0fl} \; = \; 1.9 \pm 0.5 \;\gev \qquad {\rm and} \qquad
a^-_{0fl} \; = \; 0.11^{+0.04}_{-0.02} \;\fm \,.
\end{equation}

For $E^+$, the energy of the axial vector state, we obtain no stable sum rule.
Although the expressions for $E^-$ and $E^+$ are equal in lowest order
perturbation theory, higher order perturbative contributions and the gluon
condensate lead to a splitting in such a way that for the same values of
$s_0$ and $T$ the resulting value for $E^-$ is higher than that for $E^+$.

%%%%%%%%%%%%%%
% Conclusions%
%%%%%%%%%%%%%%

\newsection{Summary and conclusions}

The analysis of the gauge invariant gluon field strength correlator
by QCD sum rule methods allows to establish a relation between the gluon 
condensate and the correlation length. In order to apply the sum rule
technique which consists in the comparison of a phenomenological Ansatz
with a theoretical expression obtained from the operator product expansion
we interpret the gluon correlator as the correlator of two colour 
neutral hybrid states composed of a (fictitious) heavy quark transforming
under the adjoint representation and the gluon field. The former serves as
the source for the gauge string in the correlator.

In this approach the decomposition in two invariant functions $D^+$ and 
$D^-$ is more appropriate than the decomposition of eq. \eqn{eq:2.2},
since $D^-$ receives only contributions from $1^-$ and $D^+$ from
$1^+$ intermediate states (ignoring the decoupled spin of the heavy
octet quark). Therefore these functions show simple exponential behaviour
at large distances and {\em not} $D$ and $D_1$. The perturbative expressions
for $D^+$ and $D^-$ are nearly degenerate, but the gluon condensate
contributes with different sign. It stabilises the sum rule for $D^-$ and
destabilises for $D^+$.

The value of the binding energy for the lowest intermediate $1^-$ state
(the inverse correlation length of the correlator) with three flavours is
determined to be $E^-_{3fl} = 1/a^-_{3fl} \approx 1.5 \pm 0.4 \,\gev$ and
with zero flavours to be $E^-_{0fl} = 1/a^-_{0fl} \approx 1.9 \pm 0.5 \,\gev$.
The main sources of uncertainty are the choice of the continuum threshold
$s_0$ and the value of the gluon condensate. 

Though we find no stable sum rule for the axial vector state we have from
the difference of the expressions for the $1^-$ and $1^+$ state strong
evidence for the counterintuitive result that the $1^+$ state is lighter
than the vector state. 

The gauge invariant gluon correlator has been calculated on the lattice using
the cooling technique \cite{gmp:97,egm:97}. There, the analysis has been made
by assuming at large distances an exponential behaviour for the invariant
functions $D$ and $D_1$, which in light of the present investigation seems
less justified than the same Ansatz for the functions $D^-$ and $D^+$. The
results of the lattice calculation are in qualitative, but not quantitative
agreement with the sum rule results. The lattice researchers find correlation
lengths for $D$ and $D_1$, $a$ and $a_1$, which are degenerate within the
errors. The computations have been done in quenched QCD and with four dynamic
flavours of staggered fermions at a bare quark mass of $d\cdot m_q = 0.01$
where $d$ denotes the lattice spacing. They found \cite{egm:97}:
\begin{eqnarray}
& & E^- \; = \; E^+ \; = \; \frac{1}{a} \; = \; 0.90\pm 0.14 \;\gev
\qquad \mbox{for 0 flavours and} \nn \\
\smvs
& & E^- \; = \; E^+ \; = \; \frac{1}{a} \; = \; 0.58\pm 0.10 \;\gev
\qquad  \mbox{for 4 flavours.}
\end{eqnarray}

A preliminary analysis of the lattice data based on an exponential behaviour
for $D^+$ and $D^-$ \cite{meg:98} leaves the values essentially unchanged but
indicates a splitting of $E^+$ and $E^-$ in the same direction as proposed
by the sum rules! The reader should also note the increase of the correlation
length from zero to four flavours which is predicted by the sum rules as well
where it is mainly due to the decrease of the gluon condensate.

In another approach \cite{bbv:98} the exponential behaviour of the functions
$D^+$ and $D^* = z^2\partial/\partial z^2 D_1$ for quenched QCD could be
extracted by analysing field insertions into a Wilson loop and assuming
factorisation as in the model of the stochastic vacuum \cite{dos:87,ds:88}.
The resulting values for the correlation lengths are smaller than those of
the direct lattice calculations \cite{bbv:98} and thus compare more
favourably with our results:
\begin{equation}
E^+ \; = \; \frac{1}{a^+} \; = \; 1.64 \,\gev
\quad \mbox{and} \quad
E^* \; = \; \frac{1}{a^*} \; = \; 1.04 \,\gev
\quad \mbox{for 0 flavours.} 
\end{equation}

The sum rule analysis shows that the state investigated here namely a gluon
confined by an octet source has a much higher energy than the corresponding
state in HQET. A similar analysis of a light quark bound by a source in the
fundamental representation \cite{bbbd:92} yielded an energy which is by a
factor 2 to 4 smaller. This is to be expected on general grounds \cite{nsvz:84}
since the case treated here is nearer to a glueball than to a heavy meson.

\newpage \noindent
{\Large\bf Acknowledgements}

\vspace{3mm} \noindent
The authors would like to thank N. Brambilla, D. Gromes, E. Meggiolaro
and A. Vairo for interesting discussions. M. Eidem\"uller thanks the
Landesgraduiertenf\"orderung at the University of Heidelberg for support
and M. Jamin would like to thank the Deutsche Forschungsgemeinschaft for
their support.

%%%%%%%%%%%%
% Appendix %
%%%%%%%%%%%%
 
\vspace{6mm} \noindent
\appendix{\noindent\Large\bf Appendix}
 
\vspace{3mm} \noindent
The theoretical expression for the perturbative correlator up to
next-to-leading order is given by:
\begin{displaymath}
\label{eq:a.1}
\cDt^\mp_{PT}(w) \; = \; (-w)^3 \Big[\, a (p_{10}^\mp+p_{11}^\mp L) +
a^2 (p_{20}^\mp+p_{21}^\mp L+p_{22}^\mp L^2) \,\Big] \,,
\end{displaymath}
where $L=\ln(-2w/\mu)$ and the $p_{ij}^\mp$ have the values
\begin{eqnarray}
p^-_{10} &=& \frac{40}{3} \nn\\
p^-_{11} &=& -\,16 \nn\\
p^-_{20} &=& \frac{2839}{9}+18 \pi^2-96 \zeta(3)+ \left(-\frac{364}{27}-
             \frac{4 \pi^2}{9}\right)n_f \nn\\
p^-_{21} &=& -\,\frac{692}{3}-16 \pi^2 + \frac{104}{9} n_f \nn\\ 
p^-_{22} &=& 44 -\frac{8}{3} n_f \nn\\
p^+_{10} &=& -\,\frac{128}{3} \nn\\ 
p^+_{11} &=& 32 \nn\\
p^+_{20} &=& -\,\frac{5684}{9}-44 \pi^2+192 \zeta(3)+ \left(\frac{848}{27}+
             \frac{8 \pi^2}{9}\right)n_f \nn\\
p^+_{21} &=& \frac{1072}{3}+32 \pi^2 -\frac{208}{9} n_f \nn\\ 
p^+_{22} &=& -\,40+\frac{16}{3} n_f \,.\nn
\end{eqnarray}

For the convenience of the reader we also give the definition of the
Borel transformation and some useful formulae:
\begin{eqnarray}
& & \wh{B}_T \; = \; \lim_{-w,n\rightarrow \infty} \frac{(-w)^{n+1}}
{\Gamma(n+1)}\left(\frac{d}{dw}\right)^n \,,\hspace{1cm}
T \; = \; \frac{-w}{n}>0 \;\mbox{fixed} \nn \\
\smvs
& & \wh{B}_T \,\frac{1}{(E-w-i\ep)^\alpha} \; = \; \frac{1}{\Gamma(\alpha)
T^{\alpha-1}}\, e^{-E/T} \,. \nn
\end{eqnarray}

Below, we have collected some formulae for the incomplete Gamma function
which are helpful for the numerical analysis of the sum rules:
\begin{eqnarray}
\Gamma(\alpha,x) & = & \int_x^\infty e^{-t} t^{\alpha-1}dt \nn \\
\smvs
\Gamma(n,x) & = & \Gamma(n)\, e^{-x} \sum_{k=0}^{n-1}\, \frac{x^k}{k!} \,,
\hspace*{1cm} n\,=\,1,2,... \nn \\
\smvs
\Gamma'(\alpha,x) & \equiv & \frac{\partial}{\partial \alpha}\,\Gamma(\alpha,x)
\nn \\
\smvs
\Gamma'(\alpha)-\Gamma'(\alpha,x) & = & \int_0^x e^{-t} t^{\alpha-1} \ln t \ dt
\nn \\
\smvs
\Gamma'(4)-\Gamma'(4,x) & = & 11-6\gamma_E-6\Gamma(0,x) \nn \\
\smvs
& & -\,e^{-x} \left(\,11+5x+x^2+\left(6+6x+3x^2+x^3\right)\ln x\,\right) \nn \\
\smvs
\Gamma'(5)-\Gamma'(5,x) & = & 50-24\gamma_E-24\Gamma(0,x) \nn \\
\smvs
& & \hspace{-15mm} -\,e^{-x} \left(\,50+26x+7x^2+x^3+\left(24+24x+12 x^2+4 x^3+
x^4\right)\ln x \,\right) \,. \nn
\end{eqnarray}

\newpage
%\bibliography{ffref}

\end{document}